\font\eightit=cmti8
\def\myr#1{\ignorespaces $^{#1}$}
\documentclass[aps,prl,showpacs,twocolumn,groupedaddress]{revtex4}  
\usepackage{graphicx}  
\usepackage{dcolumn}   
\usepackage{bm}        
\usepackage{amssymb}   
\newcommand{\ppbar}{$p\bar p$}
\newcommand{\met}{$E_T\hspace{-1.1em}/$\hspace{0.7em}}
\newcommand{\metb}{$E_T\hspace{-1.1em}/$\hspace{1.0em}}
\newcommand{\metc}{E_T\hspace{-1.1em}/\hspace{0.6em}}

\begin{document}
\hyphenation{simu-la-tion}
\title{\begin{boldmath}Search for Scalar Leptoquark Pairs Decaying to 
$\nu \bar \nu q \bar q$ 
in \ppbar  ~Collisions at $\sqrt {s} = 1.96$~TeV\end{boldmath}}
\author{
\hfilneg
\begin{sloppypar}
\noindent
D.~Acosta,\myr {16} J.~Adelman,\myr {12} T.~Affolder,\myr 9 T.~Akimoto,\myr {54}
M.G.~Albrow,\myr {15} D.~Ambrose,\myr {43} S.~Amerio,\myr {42}  
D.~Amidei,\myr {33} A.~Anastassov,\myr {50} K.~Anikeev,\myr {15}
% ...
\end{sloppypar}
\begin{center}
(CDF Collaboration)
\end{center}
}
\affiliation{
\begin{center}
\myr 1  {\eightit Institute of Physics, Academia Sinica, Taipei, Taiwan 11529, 
Republic of China} \\
\myr 2  {\eightit Argonne National Laboratory, Argonne, Illinois 60439} \\
\myr 3  {\eightit Institut de Fisica d'Altes Energies, Universitat Autonoma
de Barcelona, E-08193, Bellaterra (Barcelona), Spain} \\
\myr 4  {\eightit Istituto Nazionale di Fisica Nucleare, University of Bologna,
I-40127 Bologna, Italy} \\
\myr 5  {\eightit Brandeis University, Waltham, Massachusetts 02254} \\
\myr 6  {\eightit University of California at Davis, Davis, California  95616} \\
\myr 7  {\eightit University of California at Los Angeles, Los 
Angeles, California  90024} \\
\myr 8  {\eightit University of California at San Diego, La Jolla, California  92093} \\ 
\myr 9  {\eightit University of California at Santa Barbara, Santa Barbara, California 
93106} \\ 
\myr {10} {\eightit Instituto de Fisica de Cantabria, CSIC-University of Cantabria, 
39005 Santander, Spain} \\
\myr {11} {\eightit Carnegie Mellon University, Pittsburgh, PA  15213} \\
\myr {12} {\eightit Enrico Fermi Institute, University of Chicago, Chicago, 
Illinois 60637} \\
\myr {13}  {\eightit Joint Institute for Nuclear Research, RU-141980 Dubna, Russia}
\\
\myr {14} {\eightit Duke University, Durham, North Carolina  27708} \\
\myr {15} {\eightit Fermi National Accelerator Laboratory, Batavia, Illinois 
60510} \\
\myr {16} {\eightit University of Florida, Gainesville, Florida  32611} \\
\myr {17} {\eightit Laboratori Nazionali di Frascati, Istituto Nazionale di Fisica
               Nucleare, I-00044 Frascati, Italy} \\
\myr {18} {\eightit University of Geneva, CH-1211 Geneva 4, Switzerland} \\
\myr {19} {\eightit Glasgow University, Glasgow G12 8QQ, United Kingdom}\\
\myr {20} {\eightit Harvard University, Cambridge, Massachusetts 02138} \\
\myr {21} {\eightit The Helsinki Group: Helsinki Institute of Physics; and Division of
High Energy Physics, Department of Physical Sciences, University of Helsinki, FIN-00044, Helsinki, Finland}\\
\myr {22} {\eightit Hiroshima University, Higashi-Hiroshima 724, Japan} \\
\myr {23} {\eightit University of Illinois, Urbana, Illinois 61801} \\
\myr {24} {\eightit The Johns Hopkins University, Baltimore, Maryland 21218} \\
\myr {25} {\eightit Institut f\"{u}r Experimentelle Kernphysik, 
Universit\"{a}t Karlsruhe, 76128 Karlsruhe, Germany} \\
\myr {26} {\eightit High Energy Accelerator Research Organization (KEK), Tsukuba, 
Ibaraki 305, Japan} \\
\myr {27} {\eightit Center for High Energy Physics: Kyungpook National
University, Taegu 702-701; Seoul National University, Seoul 151-742; and
SungKyunKwan University, Suwon 440-746; Korea} \\
\myr {28} {\eightit Ernest Orlando Lawrence Berkeley National Laboratory, 
Berkeley, California 94720} \\
\myr {29} {\eightit University of Liverpool, Liverpool L69 7ZE, United Kingdom} \\
\myr {30} {\eightit University College London, London WC1E 6BT, United Kingdom} \\
\myr {31} {\eightit Massachusetts Institute of Technology, Cambridge,
Massachusetts  02139} \\   
\myr {32} {\eightit Institute of Particle Physics: McGill University,
Montr\'{e}al, Canada H3A~2T8; and University of Toronto, Toronto, Canada
M5S~1A7} \\
\myr {33} {\eightit University of Michigan, Ann Arbor, Michigan 48109} \\
\myr {34} {\eightit Michigan State University, East Lansing, Michigan  48824} \\
\myr {35} {\eightit Institution for Theoretical and Experimental Physics, ITEP,
Moscow 117259, Russia} \\
\myr {36} {\eightit University of New Mexico, Albuquerque, New Mexico 87131} \\
\myr {37} {\eightit Northwestern University, Evanston, Illinois  60208} \\
\myr {38} {\eightit The Ohio State University, Columbus, Ohio  43210} \\  
\myr {39} {\eightit Okayama University, Okayama 700-8530, Japan}\\  
\myr {40} {\eightit Osaka City University, Osaka 588, Japan} \\
\myr {41} {\eightit University of Oxford, Oxford OX1 3RH, United Kingdom} \\
\myr {42} {\eightit University of Padova, Istituto Nazionale di Fisica 
          Nucleare, Sezione di Padova-Trento, I-35131 Padova, Italy} \\
\myr {43} {\eightit University of Pennsylvania, Philadelphia, 
        Pennsylvania 19104} \\   
\myr {44} {\eightit Istituto Nazionale di Fisica Nucleare, University and Scuola
               Normale Superiore of Pisa, I-56100 Pisa, Italy} \\
\myr {45} {\eightit University of Pittsburgh, Pittsburgh, Pennsylvania 15260} \\
\myr {46} {\eightit Purdue University, West Lafayette, Indiana 47907} \\
\myr {47} {\eightit University of Rochester, Rochester, New York 14627} \\
\myr {48} {\eightit The Rockefeller University, New York, New York 10021} \\
\myr {49} {\eightit Istituto Nazionale di Fisica Nucleare, Sezione di Roma 1,
University di Roma ``La Sapienza," I-00185 Roma, Italy}\\
\myr {50} {\eightit Rutgers University, Piscataway, New Jersey 08855} \\
\myr {51} {\eightit Texas A\&M University, College Station, Texas 77843} \\
\myr {52} {\eightit Texas Tech University, Lubbock, Texas 79409} \\
\myr {53} {\eightit Istituto Nazionale di Fisica Nucleare, University of Trieste/\
Udine, Italy} \\
\myr {54} {\eightit University of Tsukuba, Tsukuba, Ibaraki 305, Japan} \\
\myr {55} {\eightit Tufts University, Medford, Massachusetts 02155} \\
\myr {56} {\eightit Waseda University, Tokyo 169, Japan} \\
\myr {57} {\eightit Wayne State University, Detroit, Michigan  48201} \\
\myr {58} {\eightit University of Wisconsin, Madison, Wisconsin 53706} \\
\myr {59} {\eightit Yale University, New Haven, Connecticut 06520} \\
\end{center}
}
\date{\today}
\begin{abstract}
We report on a search for the pair production of
scalar leptoquarks, $LQ$, using $191$~pb$^{-1}$ of proton-antiproton collision data
recorded by the CDF 
experiment during  Run II of the Tevatron.  The leptoquarks are sought
via their decay into a neutrino and quark yielding  missing
transverse energy and several jets of  large transverse energy. 
No evidence for
leptoquark production is observed, and limits are set on 
$\sigma(p\bar p\rightarrow LQ\overline{LQ} X \rightarrow \nu\bar\nu q\bar
q X)$. Using a next-to-leading order theoretical prediction of the
cross section 
for scalar leptoquark production, we exclude first-generation leptoquarks in 
the mass interval 78 to 117
GeV$/c^2$ at the 95\% confidence level for BR($LQ\rightarrow \nu q)=100\%$.
\end{abstract} 
\pacs{12.60.-i, 13.85.Rm, 14.80.-j}
\maketitle

The remarkable symmetry between quarks and leptons in the standard model (SM) suggests
 that some more fundamental theory
may exist, which allows interactions between them. Such interactions are mediated
 by a new type of particle, a leptoquark~\cite{BRW}, which carries both lepton and baryon number. 
A leptoquark is a color-triplet boson with spin 0 or 1, and has fractional electric charge. 
Leptoquarks are predicted in many extensions of
the SM (e.g. grand unification, technicolor, and
supersymmetry with $R$-parity violation).
The Yukawa coupling of the leptoquark to a lepton and quark and
the branching ratio to a
charged lepton, denoted by $\beta$, are model dependent. 
Usually it is assumed that leptoquarks couple to only one generation to accommodate 
experimental constraints on
flavor-changing neutral currents~\cite{barion}, which allows one to
classify leptoquarks as first-, second-, or third-generation.
In $p\bar p$ collisions, leptoquarks can be produced in pairs via the strong interaction 
through $gg$ fusion or 
$q\bar q $ annihilation. 
The production rate for scalar
leptoquarks is essentially model-independent and is
determined by the known QCD couplings and leptoquark mass.

We report on a search for pair production of scalar leptoquarks, with $LQ$ decaying to
 $\nu q$,
resulting
in a jets and missing transverse energy (\met) topology.
 We use $191\pm11$~pb$^{-1}$~\cite{lum}
  of $p \bar p$ collision data at a center-of-mass energy  of 1.96~TeV
recorded by the Collider Detector at Fermilab (CDF) during the
Tevatron Run II.
This analysis 
is sensitive to leptoquarks of all three generations with $\beta\approx 0$. 
The previous lower mass limit of 98~GeV$/c^2$~\cite{d0lq1} on 
first-generation leptoquarks in this final state
was set by the D\O~Collaboration. The CDF Collaboration has also published
\cite{cdflq23} lower mass limits 
of 123~GeV$/c^2$ and 148~GeV$/c^2$ respectively on second- and third-generation
leptoquarks in the \met  ~plus  heavy-flavor jets final state. 
Limits on leptoquark production from the Tevatron Run I and HERA experiments as of 1999 
are summarized in \cite{review}.  

CDF is  a general-purpose detector 
 that is described in detail elsewhere~\cite{cdf}. The components relevant
 to this analysis are briefly described here.
The charged-particle tracking system is closest to the beam pipe,
and consists of multi-layer silicon detectors and a large open-cell drift chamber
covering the pseudorapidity~\cite{eta} region $|\eta|< 1$.
The tracking  system  is enclosed in a superconducting solenoid,
which in turn is surrounded by a calorimeter. The CDF calorimeter system is organized into electromagnetic
 and hadronic sections segmented in projective tower geometry, and covers  the region $|\eta|< 3.6$. 
The electromagnetic calorimeters utilize a lead-scintillator sampling  technique, whereas
the hadron calorimeters use iron-scintillator technology.
 The central muon-detection system, used for this analysis,
is located outside of the
calorimeter and covers the range $|\eta|<1$.

This search centers on selecting events with large
\metb and a pair of jets that are acollinear in the transverse plane, because of the neutrinos in the final state.
The \metb~\cite{eta} is defined as the energy imbalance in the plane transverse
to the beam direction.
A jet is defined as a localized energy deposition in the calorimeter
 and is reconstructed using a  
cone 
algorithm with fixed radius $\Delta R \equiv\sqrt{\Delta\eta^2+\Delta\phi^2}=0.4$
 in $\eta-\phi$ space~\cite{jetcorr}.
We correct~\cite{jetcorr} jet $E_T$ measurements and \metb for detector effects. 

The data sample for this analysis was collected using an inclusive
\metb trigger, which is distributed across three levels of online event selection. 
In the first and second levels of the trigger, \metb is required  to be
greater than $25$~GeV and is calculated by summing over calorimeter 
trigger towers~\cite{trigger} with transverse energies
above 1~GeV. 
At Level-3 \metb is required  to be
greater than $45$~GeV and is recalculated using full calorimeter
segmentation with
a tower energy threshold of  100~MeV. 
 We use events from the inclusive high-$p_T$ lepton ($e$ or $\mu$) samples
 to measure the trigger efficiency directly from data.
To reduce systematic effects associated with the online trigger threshold,
we select events offline with $\metc>60$~GeV, where the trigger is fully efficient.

The event electromagnetic fraction ($F_{em}$) and charged fraction
($F_{ch}$)~\cite{clean} are used to remove events associated with beam halo
and cosmic ray sources. We reject events that contain 
little energy in the electromagnetic section of the calorimeter or 
that have mostly neutral-particle jets, by requiring $F_{em}>0.1$ and $F_{ch}>0.1$. 
There are  148,462 events
in our analysis sample after the initial selection.

The dominant backgrounds to the leptoquark search in the jets and
\metb signature are QCD multi-jet production, $W$ and $Z$ boson
production in association with one or more
jets, and  top quark pair production.
The {\sc alpgen} generator \cite{alpgen} was used for the simulation of the $W$ and $Z$
boson plus parton production, with  {\sc herwig}~\cite{herwig} used to model parton 
showers.  We use the exclusive  $Z(\rightarrow ee)+1$ jet sample to determine a scale factor between 
data and simulation, and apply this factor to all  $W/Z+$jets simulation samples.
{\sc herwig} was also used to estimate the contribution from $t\bar t$ production.

Data selection requirements were chosen to maximize the statistical significance of the leptoquark signal over background
events based on studies of simulated event samples before the signal
region data were examined. In addition to $\metc>60$~GeV,
the signal region  is defined by requiring that the two highest 
$E_T$ jets ($E_{T}^{j_1}> 40$~GeV, $E_{T}^{j_2}> 25$~GeV) 
be in the central region $|\eta|<1$.
A third jet with  $E_T> 15$~GeV and  $|\eta|<2.5$ is allowed, and we
veto events with any  additional jets
with  $E_T> 15$~GeV and  $|\eta|<3.6$. 
To reject events with \metb resulting from jet energy
mismeasurement, we require that the opening angle in
the transverse  plane between the two highest $E_T$ jets satisfy
$80^{\circ}<\Delta\phi(j_1,j_2)<165^{\circ}$.
The \metb  direction must not be parallel
to any of the jets; we require the minimum azimuthal separation between
 the direction of the jets and \metb to satisfy $30^{\circ}< \min\Delta\phi(j,\metc)<135^{\circ}$.
 The \metb also must not be antiparallel to the leading $E_T$ jet: 
$100^{\circ}<\Delta\phi(j_1,\metc)<165^{\circ}$. These criteria reject most of the 
QCD multi-jet background events. To reduce the background 
contribution from $W/Z+$jets and $t\bar t$ production, 
we reject events with one or more identified leptons with $E_T >10$~GeV 
(electron candidates) or $p_T>10$~GeV$/c$ (muon candidates). 
Criteria similar to those in \cite{lepid} are used to identify the leptons.
To further reduce this background  we require each jet not to be highly electromagnetic 
(jet electromagnetic fraction $<0.9$) and to have 4 or more associated tracks for central jets ($|\eta|<1$).

Two methods are employed to estimate the QCD multi-jet contribution in the
signal region directly from the inclusive \metb data sample. Among all the
offline analysis selection requirements,  the azimuthal
angular separation requirement between the \metb direction and a jet is
most effective at removing QCD multi-jet events. 
Therefore, for the first method, in addition to the signal region we define a region
which is rich in QCD multi-jet events by requiring that a jet
is close to the \metb direction 
($20^{\circ}<\min\Delta\phi(j,\metc)<27^{\circ}$).
Studies of simulated QCD multi-jet samples show that the shape of the \metb
distribution in this region is similar to the \metb distribution in the
signal region.
We use \metb and $\min\Delta\phi(j,\metc)$ requirements to define
 four kinematic regions:
\begin{itemize}
\item[A)] $50<\metc<57$~GeV,
$20^{\circ}<\min\Delta\phi(j,\metc)<27^{\circ}$.
\item[B)] $\metc>60$~GeV, $20^{\circ}<\min\Delta\phi(j,\metc)<27^{\circ}$.
\item[C)] $50<\metc<57$~GeV, $30^{\circ}<\min\Delta\phi(j,\metc)<135^{\circ}$.
\item[D)] $\metc>60$~GeV, $30^{\circ}<\min\Delta\phi(j,\metc)<135^{\circ}$.
\end{itemize}
 The regions A, B and C are used to extrapolate the QCD multi-jet contribution into the signal region D:
$ N_D = \frac{N_B}{N_A}N_C,$ where
$N_A$, $N_B$, and $N_C$ are the remaining number of events in regions A, B, and
C, after the $W/Z+$jets and $t\bar t$ contributions have been subtracted.
 For the second method, the combined selection requirement efficiency
 is measured as a function of \metb in an independent inclusive jet sample at low \met.
The extrapolated results of this measurement is then applied to the inclusive
\metb sample after the $W/Z+$jets and $t\bar t$ contributions have been subtracted.
We predict $15.0\pm8.0$ and $21.5\pm12.4$ multi-jet events for the first and second methods respectively.
We take the weighted average and uncertainty  of the two methods as our estimate
 of the multi-jet background.

We check the simulation predictions for $W/Z+$jets with data in a control region,
which is defined by requiring, in addition to 2 or 3 jets, $\metc>60$~GeV
 and at least one electron or muon. We observe 144 events in our inclusive  
\metb sample, which is in excellent agreement with $154.3\pm27.9$ 
events predicted from SM processes. 

The total detection efficiency ($\epsilon_{LQ_1}$) for the first-generation scalar leptoquark 
($LQ_1$) signal is estimated using
the {\sc pythia} event generator~\cite{pythia},
 and the CDF detector simulation program. The
{\sc pythia} underlying event simulation was tuned to reproduce CDF data \cite{field}. The samples were generated 
using the CTEQ5L~\cite{cteq5l} 
parton distribution functions (PDF), with the renormalization and factorization scales set to $\mu=m_{LQ_1}$.
Table \ref{tab:sigacc} lists the total detection efficiency $\epsilon_{LQ_1}$ and 
 the corresponding total fractional uncertainty $\delta_{tot}$
 for various leptoquark masses. Also listed are the NLO cross sections~\cite{NLO} calculated
 for two choices of the $\mu$ scale.
\begin{table}
  \caption{Summary of the first-generation scalar leptoquark detection efficiency ($\epsilon_{LQ_1}$), 
the relative uncertainty on detection efficiency ($\delta_{\rm tot}$), and the 
next-to-leading order cross section
 ($\sigma_{\rm NLO}$) for two choices of the renormalization scale as functions of leptoquark mass.
  }
  \label{tab:sigacc}
  \begin{center}
    \begin{tabular}{ccccc} \hline\hline
&&&\multicolumn{2}{c}{$\sigma_{\rm NLO}$(pb)}\\
$m_{LQ_1}$ (GeV$/c^2$) & $\epsilon_{LQ_1}$ & $\delta_{\rm tot}$ (\%) &
$\mu=m_{LQ_1}$ &$\mu=2m_{LQ_1}$ \\\hline
  75            & 0.0073 & 29 & 69.4 & 58.8\\ 
  80            & 0.0113 & 26 & 49.2&41.5 \\ 
  90            & 0.0187 & 23 & 26.0&22.1 \\ 
 100            & 0.0300 & 20 & 14.6& 12.5\\ 
 110            & 0.0431 & 16 & 8.4& 7.4 \\ 
 115            & 0.0482 & 15 & 6.7&5.8  \\ 
 125            & 0.0590 & 15 & 4.2&3.6  \\ 
 150            & 0.0828 & 13 & 1.4 &1.3 \\ 
175            & 0.1010 & 12 & 0.57&0.51  \\ \hline\hline
    \end{tabular}
  \end{center}
\end{table}
 The systematic uncertainty on the signal acceptance
 includes the uncertainties due to modeling gluon radiation from the initial-state 
or final-state partons (10\%),
and the choice of the PDF (4\%). The limited size of the leptoquark simulation
samples gives a 3\% statistical uncertainty.
The signal acceptance uncertainty due to the jet energy scale varies from
  4\% to 26\%, and the uncertainty on
 the luminosity is 6\%. The
uncertainty on the trigger efficiency is 1\%.
The theoretical uncertainties on the
renormalization and factorization scales are not included here, since we
conservatively choose the NLO cross section setting $\mu = 2 m_{LQ_1}$ to extract 
the limits on leptoquark mass. 
This choice of scale is found to reduce the cross section prediction by 15\% relative to 
$\mu = m_{LQ_1}$ ~\cite{NLO}.

\begin{figure}
\centerline{
\resizebox{8cm}{!}{\includegraphics{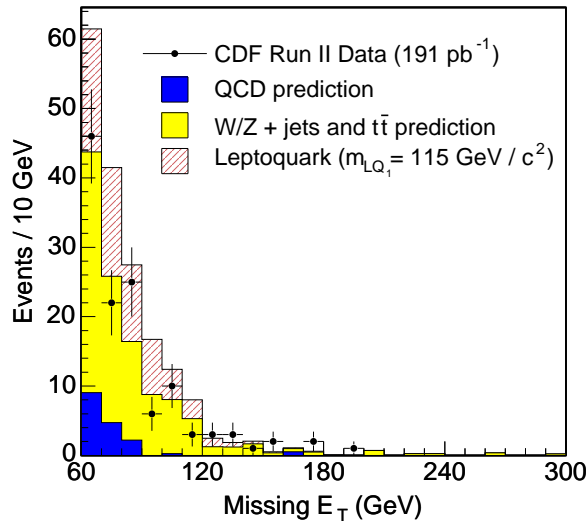}}
}
\caption
{The \metb distribution in the leptoquark signal region for data
(solid points) compared to SM background (shaded
histograms).
Also shown is the expected distribution arising from
leptoquark production and decay at a mass of 115~GeV$/c^2$
(hatched histogram).
}
\label{fig:met}
\end{figure}

\begin{figure}
\centerline{
  \resizebox{8cm}{!}{\includegraphics{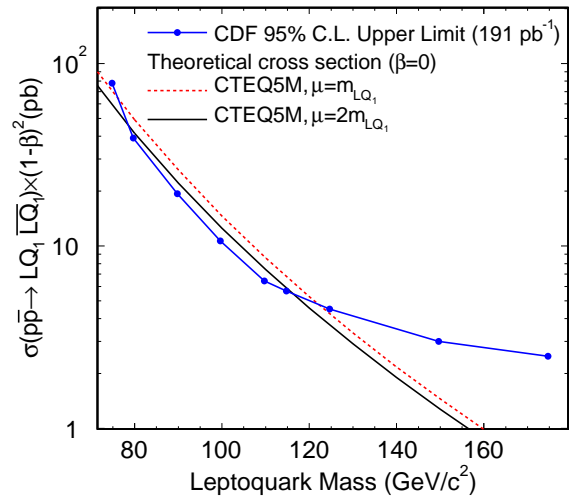}}
}
\caption{
The upper limit on the cross section times squared branching
ratio for scalar leptoquark production in the jets and \metb topology. Also shown is the
NLO cross section for $\beta=0$ for 2 choices of the
factorization/renormalization scale: $\mu=m_{LQ_1}$,  $\mu=2~m_{LQ_1}$.
}
\label{fig:limit}
\end{figure}

In the signal region, we expect $118.5\pm14.5$ events from  SM
 processes and observe 124 events.
The predicted backgrounds from SM processes are summarized
 in Table \ref{tab:sm}. 
\begin{table}[bthp]
  \caption[Number of expected events from various SM
sources in the leptoquark signal region.]{The number of expected events from various SM
sources in the leptoquark signal region. The first uncertainty is from the limited simulation statistics
 and the second is from the various systematics.
}
  \label{tab:sm}
 \begin{center}
   \begin{tabular}{lc}  \hline\hline
Source & Events expected \\ \hline
$W(\rightarrow e\nu) + $jets     & $6.1   \pm 1.4  \pm 1.5$  \\
$W(\rightarrow \mu\nu) + $jets   & $21.7  \pm 2.3  \pm 2.8$  \\
$W(\rightarrow \tau\nu) + $jets  & $28.4  \pm 3.8  \pm 4.1$  \\
$Z(\rightarrow \mu\mu) + $jets   & $1.1   \pm 0.2  \pm 0.2$ \\
$Z(\rightarrow \tau\tau) + $jets & $0.9   \pm 0.2  \pm 0.2$ \\
$Z(\rightarrow \nu\nu) + $jets   & $39.1  \pm 2.8  \pm 3.6$\\
$t\bar t$                        & $4.3   \pm 0.4  \pm 0.3$ \\  
QCD                              & $16.9  \pm 6.7$            \\
Total Events                   & $118.5 \pm 14.5$  \\ \hline\hline
    \end{tabular}
  \end{center}
\end{table}
In Figure~\ref{fig:met} the predicted \metb distribution
is compared with the distribution observed in data.
 No evidence for
leptoquark production is observed.
We calculate  the upper limit 
 at the 95\% confidence level (C.L.) on the pair production cross section times the square of the branching ratio
  of the leptoquark to a quark and a neutrino using a Bayesian approach~\cite{cutoff} with a flat prior 
for the signal cross section and Gaussian priors for 
acceptance and background uncertainties. 
The upper limit on the cross section times $(1-\beta)^2$ is shown in Figure \ref{fig:limit} 
and is compared with the theoretical cross sections. 
The theoretical cross sections for scalar
 leptoquark production have been calculated at NLO using CTEQ5M~\cite{cteq5l} PDFs.

In conclusion, we performed a search for leptoquarks in the jets and
\metb topology using 191~pb$^{-1}$ of CDF Run II
data. No evidence for leptoquarks is observed. We set an upper limit 
on the production cross section at the 95\% C.L.  Assuming a leptoquark
 decays into a neutrino and quark with 100\% branching ratio, we exclude the
 mass interval from 78 to 117~GeV$/c^2$ for first-generation scalar leptoquarks.
 This extends  the previous limit for the first-generation scalar leptoquark of 98~GeV$/c^2$~\cite{d0lq1}.

We thank the Fermilab staff and the technical staffs of the participating institutions
 for their vital contributions. This work was supported by the U.S. Department of
 Energy and National Science Foundation; the Italian Istituto Nazionale di Fisica Nucleare; 
the Ministry of Education, Culture, Sports, Science and Technology of Japan; the Natural
 Sciences and Engineering Research Council of Canada; the National Science Council of the
 Republic of China; the Swiss National Science Foundation; the A.P. Sloan Foundation;
 the Bundesministerium fu\"r Bildung und Forschung, Germany; the Korean Science and Engineering 
Foundation and the Korean Research Foundation; the Particle Physics and Astronomy Research Council 
and the Royal Society, UK; the Russian Foundation for Basic Research; the Comision Interministerial
 de Ciencia y Tecnologia, Spain; in part by the European Community's Human Potential Programme
 under contract HPRN-CT-20002, Probe for New Physics.


\begin{thebibliography}{}
\bibitem{eta} CDF uses a cylindrical coordinate system in which $\theta$ 
is  the polar angle to the proton beam, $\phi$ is the azimuthal angle about the beam axis, and 
pseudorapidity is defined as  $\eta=-$ln tan($\theta$/2). The transverse energy and transverse momentum are defined 
as $E_T$ = $E$sin$\theta$ and $p_T$ = $p$sin$\theta$, where $E$ is energy measured in the calorimeter and $p$
 is momentum measured by the tracking system. The missing transverse energy vector, \met, 
is $-\sum_i E^i_T{\bm n}_i$, where ${\bm n}_i$ is the unit vector in the azimuthal plane that points 
from the beamline to the $i$th calorimeter tower.
\bibitem{trigger} The physical calorimeter towers are organized into larger trigger towers, covering approximately
0.26 in $\Delta\phi$ and $0.22$ in $\Delta\eta$.
\end{thebibliography}
\end{document}